\documentclass[letterpaper]{article} %
\usepackage{aaai24}  %
\usepackage{times}  %
\usepackage{helvet}  %
\usepackage{courier}  %
\usepackage[hyphens]{url}  %
\usepackage{graphicx} %
\usepackage{natbib}  %
\usepackage{caption} %
\usepackage{algorithm}
\usepackage{algorithmic}
\usepackage{amsmath,amssymb,amsfonts,mathtools,amscd}
\usepackage{newfloat}
\usepackage{listings}
\DeclareCaptionStyle{ruled}{labelfont=normalfont,labelsep=colon,strut=off} %
\floatstyle{ruled}
\newfloat{listing}{tb}{lst}{}
\floatname{listing}{Listing}

\title{DrFuse: Learning Disentangled Representation for Clinical Multi-Modal Fusion with Missing Modality and Modal Inconsistency}
\author{
    Wenfang Yao\equalcontrib\textsuperscript{\rm 1},
    Kejing Yin\equalcontrib\textsuperscript{\rm 2},
    William K. Cheung\textsuperscript{\rm 2},
    Jia Liu\textsuperscript{\rm 3},
    Jing Qin\textsuperscript{\rm 1}
}
\affiliations{
    \textsuperscript{\rm 1}School of Nursing, The Hong Kong Polytechnic University\\
    \textsuperscript{\rm 2}Department of Computer Science, Hong Kong Baptist University\\
    \textsuperscript{\rm 3}Shenzhen Institute of Advanced Technology, Chinese Academy of Sciences\\

    dorothy.yao@polyu.edu.hk, \{cskjyin, william\}@comp.hkbu.edu.hk, jia.liu@siat.ac.cn, harry.qin@polyu.edu.hk
}

\usepackage{bibentry}

\usepackage{booktabs}
\usepackage{amsfonts,amsthm,amssymb,amsmath}
\usepackage{dsfont}
\usepackage[table,xcdraw]{xcolor}

\theoremstyle{definition}
\newtheorem{definition}{Definition}

\usepackage{xspace}
\newcommand{\drfuse}{\texttt{DrFuse}\xspace}
\newcommand{\ie}{\textit{i.e.}}

\usepackage{subcaption}
\usepackage{multirow}

\begin{document}

\maketitle

\begin{abstract}
The combination of electronic health records (EHR) and medical images is crucial for clinicians in making diagnoses and forecasting prognosis. Strategically fusing these two data modalities has great potential to improve the accuracy of machine learning models in clinical prediction tasks. However, the asynchronous and complementary nature of EHR and medical images presents unique challenges. Missing modalities due to clinical and administrative factors are inevitable in practice, and the significance of each data modality varies depending on the patient and the prediction target, resulting in inconsistent predictions and suboptimal model performance. To address these challenges, we propose \drfuse to achieve effective clinical multi-modal fusion. It tackles the missing modality issue by disentangling the features shared across modalities and those unique within each modality. Furthermore, we address the modal inconsistency issue via a disease-wise attention layer that produces the patient- and disease-wise weighting for each modality to make the final prediction. We validate the proposed method using real-world large-scale datasets, MIMIC-IV and MIMIC-CXR. Experimental results show that the proposed method significantly outperforms the state-of-the-art models. Our implementation is
publicly available at \url{https://github.com/dorothy-yao/drfuse}.
\end{abstract}

\section{Introduction} \label{sect:intro}
Clinicians rely on data from various sources, including electronic health records (EHR) and medical imaging, to make diagnosis and forecast prognosis~\cite{aljondi2020diagnostic}. For instance, when diagnosing pneumonia, EHR data like blood tests provides information about the patient's infection status and immune response, while medical images like Chest X-ray (CXR) can reveal the extent of inflammation in the lungs~\cite{hoare2006pneumonia}. Integrating these data modalities could shed light on a more comprehensive and accurate understanding of the patient's health condition, potentially leading to a better clinical outcome \cite{huang2020fusion}. With the increasing availability of digital clinical data, research efforts have recently been made to employ multi-modal machine learning approaches to improve the performance of clinical prediction tasks, including disease prediction~\cite{mlhc2022hayatmedfuse} and mortality prediction~\cite{lin2021empirical}.

The multi-modal data fusion, \ie, the process of combining different data modalities, plays a central role in the effective utilization of multi-modal clinical data. %
Despite the recent effort, their applications to real-world data are still hindered due to the complex and complementary nature of multi-modal clinical data. Specifically, there are fundamentally challenging issues need to be addressed:

\vspace{0.25em}\noindent\textbf{Challenge 1: Missing modality in a highly heterogeneous setting.} Many existing work on clinical multi-modal learning assumes that both EHR and medical images are available for all training and testing samples, which is not practical in real-world clinical settings. For instance, the MIMIC-IV~\cite{johnson2023mimic}, a real-world ICU dataset, has less than 20\% of patients with X-ray images. Many in-hospital patients requiring X-ray scans cannot undergo the procedure due to clinical or administrative reasons, resulting in a significant number of patients with missing modalities~\cite{huang2020fusion}. Similar problems exist in other domains like tumor segmentation on multi-modal MRI images~\cite{zhao2022modality}, where generative machine learning models are commonly used to synthesize the missing modality~\cite{sharma2019missing}. However, accurately generating missing medical image using EHR data is infeasible because EHR contains information about a patient's clinical conditions, medical history, and treatments, but they do not provide a detailed enough picture of the patient's anatomy to generate a missing modality of medical imaging, such as a chest X-ray. Late fusion is a common approach of tackling missing modality in the fusion of EHR and medical imaging, where separate prediction models are learned for different modality and the fusion happens only in the decision-level~\cite{huang2020fusion}. This approach fails to fully utilize the interactions between modalities, leading to undesirable suboptimal performance. Therefore, effectively capturing the complex interactions between highly heterogeneous modalities while handling missing modality remains an open challenge.

\vspace{0.25em}\noindent\textbf{Challenge 2: Modal inconsistency and patient-specific modal significance.} Even with fully observed data modalities, inconsistencies can arise when different modalities, such as EHR and CXR, provide inconsistent or even contradictory information regarding the prediction targets. For example, in mortality prediction, patients with meningitis may be identified as having a high risk of mortality based on EHR data due to the severity of their symptoms, while their CXR may not show any signs of complications~\cite{brouwer2010epidemiology}. Conversely, for patients with pneumothorax, CXR may predict a high risk of mortality while EHR may not indicate high mortality due to the non-specific nature of the symptoms~\cite{zarogoulidis2014pneumothorax}. The patient variation makes it even more challenging as the significance of different data modal depends on the patient's medical condition. For example, diabetic patients without specific symptoms or conditions are usually not recommended to take X-rays, while those who develop complications like foot or dental problems need X-rays to assist in diagnosing and treatment planning~\cite{ahmad2016diabetic}. Without appropriately account for such inconsistency and patient-specific significance between modalities, the accuracy of model prediction could be greatly compromised, leading to suboptimal clinical outcomes. How to effectively handle modal inconsistency and patient-dependent modal significance in multi-modal learning remains an unresolved research problem.

\vspace{0.25em}
To address the above challenges, we propose a novel method: \textit{Learning \underline{D}isentangled \underline{R}epresentation for Clinical Multi-Modal \underline{Fus}ion} (\drfuse).
We hypothesize that EHR and medical images share a common information component. To leverage this shared information, our core idea is to disentangle the shared information from the modality-distinct information of EHR and medical images. By doing so, we learn a shared representation that captures the common information across both modalities, which enables us to make more accurate predictions even when one modality is unavailable, as the shared information can be inferred from the available modality. To further utilize distinct information from each modality and allow the patient-dependent modal significance to be captured, we propose a disease-aware attention fusion module which is regulated by a novel attention weight ranking loss.
To summarize, our main contributions are three-fold:
\begin{itemize}
    \item We propose \drfuse to fully utilize information shared across modalities with disentangled representation learning. It tackles the missing modality issue as the shared information is still preserved with the available modality robustly under an end-to-end learning paradigm.
    \item \drfuse captures the patient-specific significance of EHR and medical images for each prediction target and therefore tackles the modal inconsistency problem. To the best of our knowledge, this is the first work addressing the modal inconsistency issue for highly heterogeneous clinical multi-modal data.
    \item Our experimental results show that \drfuse significantly outperforms state-of-the-art models on the phenotype classification task in the real-world large-scale MIMIC-IV dataset.
\end{itemize}

\begin{figure*}
    \centering
    \begin{subfigure}[b]{0.65\textwidth}
         \centering
         \includegraphics[width=0.98\linewidth]{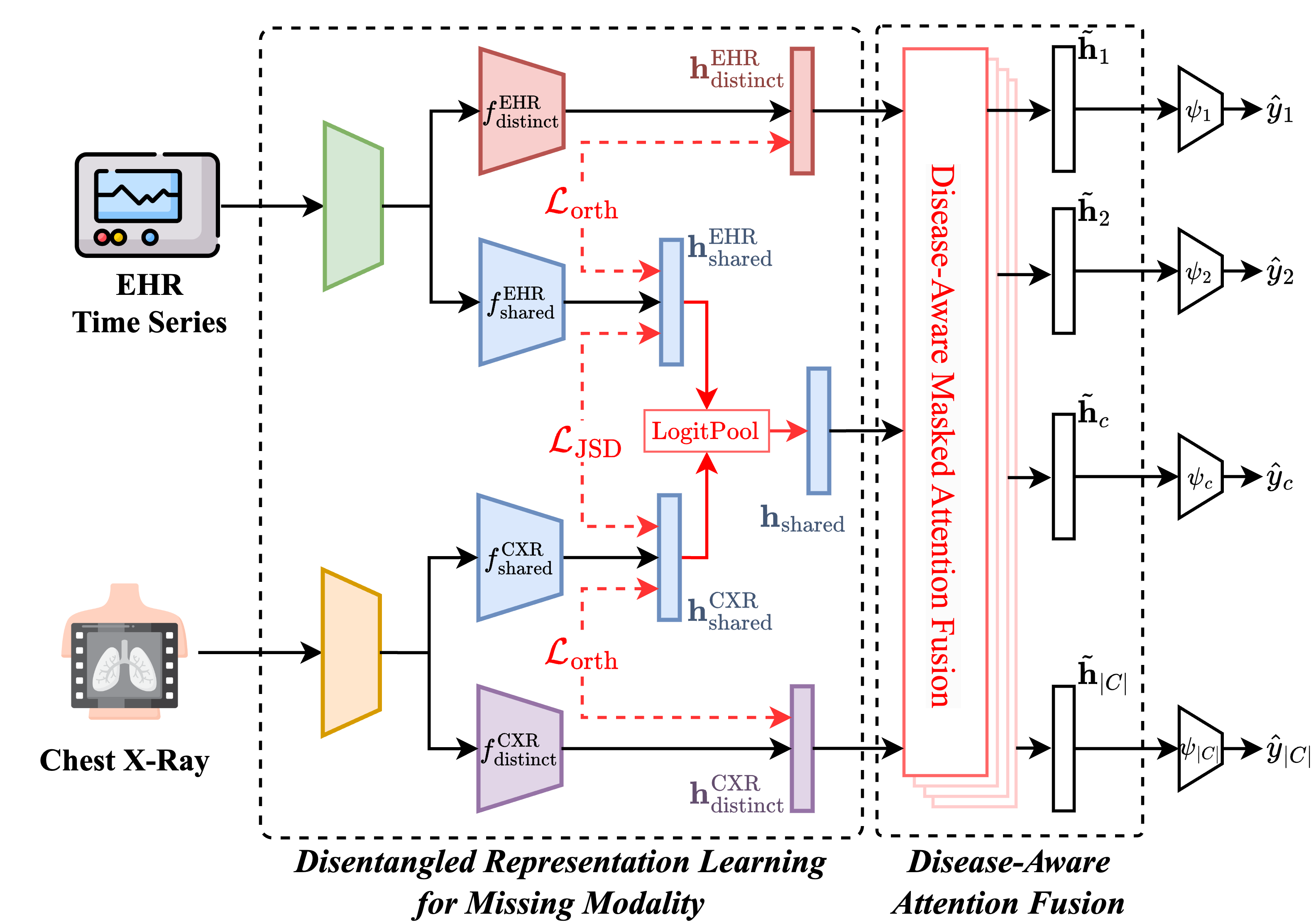}
         \caption{The architecture overview of \drfuse.}
         \label{fig:overview_full_modality}
     \end{subfigure}
     \hfill
     \begin{subfigure}[b]{0.34\textwidth}
         \centering
         \includegraphics[width=\linewidth]{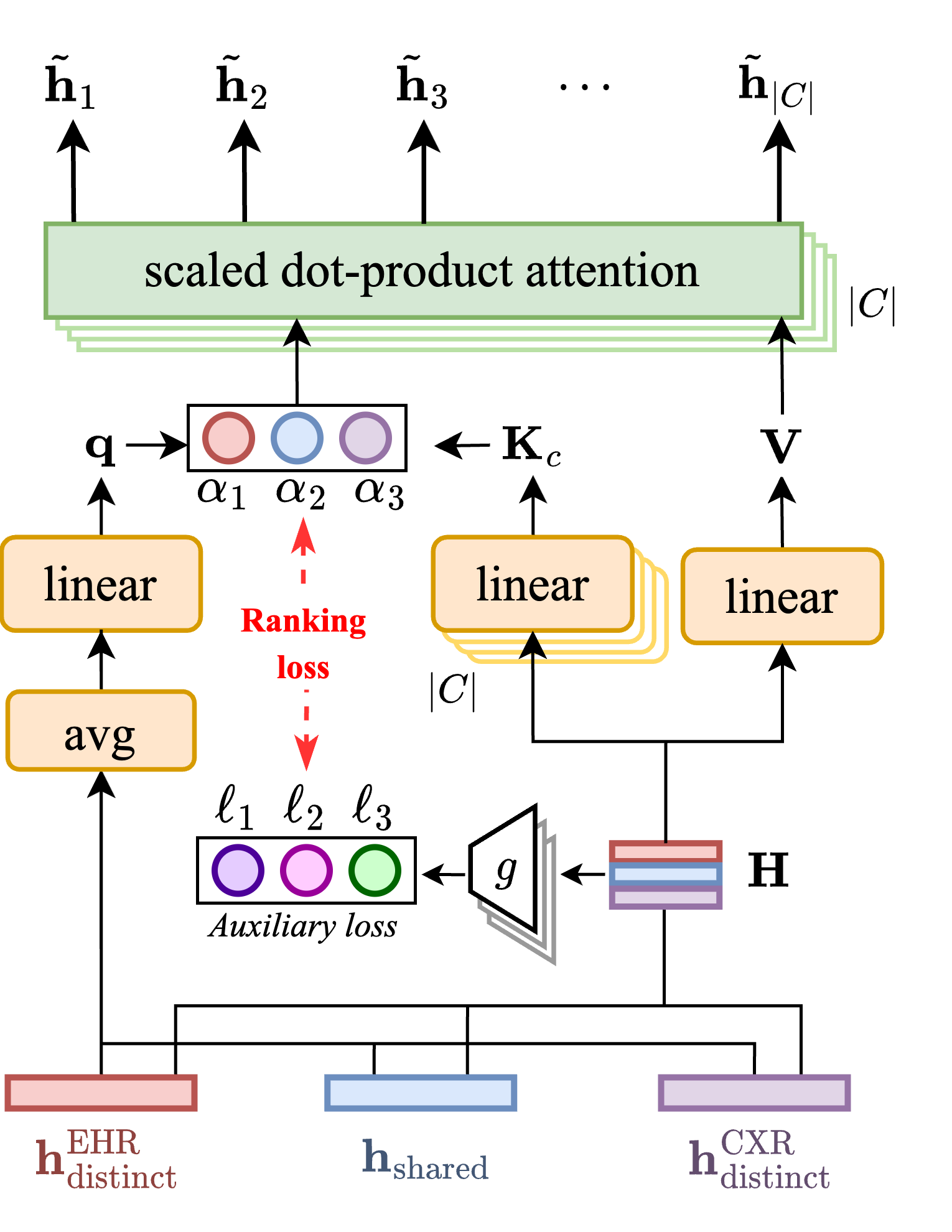}
         \caption{The disease-aware attention fusion module.}
         \label{fig:attention_module}
     \end{subfigure}
    \caption{The overview of the proposed model, \drfuse. It consists of two major components. Subfigure (a): A shared representation and a distinct representation are learned from EHR and CXR, where the shared ones are aligned by minimizing the Jensen–Shannon divergence (JSD). A novel logit pooling is proposed to fuse the shared representations. Subfigure (b): The \textit{disease-aware attention fusion} module captures the patient-specific modal significance for different prediction targets by minimizing a ranking loss.}\vspace{-1em}
    \label{fig:overview}
\end{figure*}

\section{Related Work}
\textbf{Multi-modal learning for healthcare.}
It has been shown that fusing multiple modalities has great potential to enhance machine learning models for clinical tasks such as prognosis prediction~\cite{kline2022multimodal}, phenotyping classification~\cite{mlhc2022hayatmedfuse} and medical image segmentation~\cite{huang2020multimodal}. Various data modalities, including electronic health record(EHR), clinical notes, Electrocardiogram(ECG), omics, chest X-rays, Magnetic Resonance Imaging (MRI), and computed tomography(CT), have been studied in the context of multi-modal learning~\cite{multimodalAD-review, EHRimagefusingreview}.
For example, \cite{daft-polsterl2021combining} combined 3D image and tabular information for diagnosis. Both \cite{huang2020multimodal} and \cite{PEdetection-zhi2022multimodal} fused CT images and EHR for Pulmonary Embolism(PE) diagnosis.

\noindent \textbf{Missing modality.} Although the available modalities are abundant, in practice, some modalities are inevitably missing \cite{huang2020fusion}. Late fusion is a common solution to handle the missing modality~\cite{yoo2019deep}. It aggregates the predictions from each modality with weighted sum or major voting. As each modality is modeled independently, the interaction across modalities cannot be fully captured and utilized~\cite{huang2020fusion}. Some recent research adopted generative methods to impute or reconstruct the missing modality on an instance or embedding level for compensation.
\cite{ma2021smil} reconstructs the features of missing modality by a Bayesian meta-learning framework. \cite{mlhc2022hayatmedfuse}
utilized an LSTM layer to generate a representative vector for general cases. \cite{zhang2022m3care} proposed to impute in the latent space with auxiliary information. These methods either require prior knowledge or assume different modalities to be similar. It has also been speculated that results relying on generating missing representation may not be robust~\cite{li2023multi}. Another method is to disentangle the shared and complementary information across modalities and used the shared information for reconstruction or downstream tasks~\cite{chen2019robust, shen2019brain, shaspec2023}. Nevertheless, most of these works focus on modalities with much shared information in common, for example, using four modalities of MRI for brain tumor segmentation. How to handle missing modality in a highly heterogeneous setting, like the fusion of EHR and medical image, remains an open challenge.

\noindent \textbf{Modal inconsistency.} The issue of model inconsistency has been recognized in different domains. For example, recent works utilize the inconsistency between image and text to detect fake news~\cite{xiong2023trimoon, sun2023inconsistent}. The modal inconsistency issue has also been investigated in sentiment analysis using text and image. However, it has not yet been discussed and addressed in the context of clinical multi-modal learning.

\section{\drfuse: The Proposed Method}

\subsection{Notations}
In this work, we focus on making clinical predictions using two modalities: electronic health records (EHR), which are recorded in the form of time series, and chest X-Ray images (CXR). We denote the EHR data of the $n^{\text{th}}$ patient by $\mathbf{X}_{(n)}^{\text{EHR}}\in\mathbb{R}^{T_n \times J}$, where $T_n$ and $J$ are the length of the time series and the number of features, respectively. We denote the CXR data by $\mathbf{X}^{\text{CXR}}_{(n)}$ and the prediction labels by $\mathbf{y}_n$. The data of patients who have both modalities is denoted by $\mathcal{D}_{\text{paired}}=\{(\mathbf{X}_{(n)}^{\text{EHR}}, \mathbf{X}^{\text{CXR}}_{(n)}, \mathbf{y}_n)\}_{n=1}^{N}$. In practice, EHR are routinely recorded in clinical process but CXR may not always be available. The data of patients who have only EHR data are denoted by $\mathcal{D}_{\text{partial}}=\{(\mathbf{X}_{n^\prime}^{\text{EHR}}, \mathbf{X}_{n^\prime}^{\text{CXR}}=\emptyset, \mathbf{y}_{n^\prime})\}_{{n^\prime}=1}^{N^\prime}$. To take full advantage of the available data, we use the joint of them as the full dataset, \textit{i.e.}, $\mathcal{D}=\mathcal{D}_{\text{paired}}\cup\mathcal{D}_{\text{partial}}$. To ease the notation, we omit the index of patient $n$ when doing so does not cause confusion.

\vspace{-0.5em}\subsection{Overview}
An overview of the proposed method is depicted in Fig.~\ref{fig:overview}. It consists of two main components. The \textit{disentangled representation learning} takes the EHR and CXR data as input and generates three representations, the EHR distinct representation $\mathbf{h}_{\text{distinct}}^{\text{EHR}}$, the CXR distinct representation $\mathbf{h}_{\text{distinct}}^{\text{CXR}}$. A novel logit pooling is proposed to generate the cross-modal shared representation $\mathbf{h}_{\text{shared}}$ while achieving effective distribution alignment between the two shared representations. To address the modal inconsistency issue, we propose a \textit{disease-aware attention-based fusion} that adaptively fuses the representations extracted in a patient- and disease-specific manner, where the modal significance for each prediction target can be respected. Finally, the \textit{channel-wise prediction} component makes prediction using the fused representation.

\subsection{Disentangled Representation Learning}
\paragraph{Modal-specific encoders.} EHR and CXR are two highly heterogeneous modalities, requiring separate models to encode the raw input data. For each modality, we employ two encoders with the same architecture to extract the shared and distinct representations with dimension of $d$.
For EHR data, we use Transformer models~\cite{vaswani2017attention} as the encoder, given by:
\begin{equation*}\small
    f^{\text{EHR}}(\mathbf{X}) = \operatorname{Transformer}\left( \left[\phi(\mathbf{x}_1)+\delta_1,\dots,\phi(\mathbf{x}_T)+\delta_T\right] \right),
\end{equation*}
where $\phi(\mathbf{x}_t)$ projects the raw EHR time series into an embedding space at time step $t$ and $\delta_t$ is the positional encoding. To extract representations from CXR, we use ResNet50~\cite{he2016deep} as the encoders for CXR data.

To reduce the number of parameters to be learned, we share the first layer in the two Transformer encoders which are expected to extract low-level features.

\paragraph{Shared representation alignment and logit pooling.}
The purpose of learning disentangled representation is to extract common information that is shared across modalities, so that this shared information can still be fully utilized even when one modality is missing. To this end, we need to align the distributions of the shared representations generated from EHR and CXR data. We interpret the shared representations $\mathbf{h}^{\text{CXR}}_{\text{distinct}}$ and $\mathbf{h}^{\text{EHR}}_{\text{distinct}}$ as logits of two probability distributions of a latent multivariate binary random variable, and minimize the Jensen–Shannon divergence (JSD) between the induced distributions $P=\sigma(\mathbf{h}^{\text{EHR}}_{\text{distinct}})$ and $Q=\sigma(\mathbf{h}^{\text{CXR}}_{\text{distinct}})$, where $\sigma(\cdot)$ denotes the standard logistic function, mapping the real-value logits $\mathbf{h}$ to a probability value.

The JSD, also known as total divergence to the average, measures the average information that each sample reveals about the source of the distribution from which it was sampled. Recent work has shown that JSD is more stable, consistent, and insensitive across a diverse range of inputs~\cite{hendrycks2019augmix}. This is particularly important as $\mathbf{h}^{\text{EHR}}_{\text{distinct}}$ and $\mathbf{h}^{\text{CXR}}_{\text{distinct}}$ are generated from encoders with very different architectures from heterogeneous input, resulting in a highly diverse range of values. Formally, the loss function of shared representation alignment is given by
\begin{equation}
    \mathcal{L}_{\text{JSD}}=\frac{1}{2} \left( \operatorname{KL}(P || M) + \operatorname{KL}(Q || M)\right),
    \label{eq:distribution_alignment}
\end{equation}
where $M=(P+Q)/2$ denotes the mixture of $P$ and $Q$, and $\operatorname{KL}$ denotes the Kullback–Leibler divergence. The logits corresponding to $M$ then can be computed by $\sigma^{-1}(M)$, where $\sigma^{-1}(\cdot)$ denotes the logit function, the inverse of the standard logistic function. We define the process of obtaining the logits of the mixture of the induced distributions from
$\mathbf{h}^{\text{EHR}}_{\text{distinct}}$ and $\mathbf{h}^{\text{CXR}}_{\text{distinct}}$ as \textit{logit pooling}, given by:
\begin{definition}[Logit Pooling]
    The logit pooling of $h_1$ and $h_2$ is given by:
    \begin{equation}\small
    \begin{aligned}
        \operatorname{LogitPool}(h_1, h_2)=&\sigma^{-1}\left( \frac{\sigma(h_1)+\sigma(h_2)}{2} \right)\\
        =& \log \frac{2e^{h_1+h_2}+e^{h_1}+e^{h_2}}{2+e^{h_1}+e^{h_2}}
    \end{aligned}
    \end{equation}
\end{definition}

Since the shared representations are aligned, when both modalities are present, we can obtain the final shared representation via the logit pooling. On the other hand, when CXR is missing, we can directly use the shared representation extracted from EHR data as the final shared representation. That is,
\begin{equation}
    \mathbf{h}_{\text{shared}}=\begin{cases}
        \operatorname{LogitPool}(\mathbf{h}_{\text{shared}}^{\text{EHR}},\mathbf{h}_{\text{shared}}^{\text{CXR}}) &\text{if $\mathbf{X}^{\text{CXR}}\not=\emptyset$,}
        \\
        \mathbf{h}_{\text{shared}}^{\text{EHR}} &\text{otherwise}.
        \end{cases}
        \label{eq:pool_shared}
\end{equation}

\paragraph{Representation disentanglement via orthogonality.} The information shared across modalities and that distinct within each modality are not naturally separated. To enable the modal-distinct ones to capture information that are not shared by the other modality, we impose orthogonality constraints to disentangle the modal-distinct information and reduce the redundancy in the shared and the modal-distinct representations~\cite{jia2020semi}. The orthogonality constraint can be enforced by minimizing the absolute value of the cosine similarities between the distinct representation and the shared representation for each modality. Formally, we have:
\begin{equation}
\begin{aligned}
    \mathcal{L}_{\text{orth}}^{\text{EHR}} =& \ell_{\text{orth}}(\mathbf{h}_{\text{shared}}^{\text{EHR}}, \mathbf{h}_{\text{distinct}}^{\text{EHR}})\\
    \mathcal{L}_{\text{orth}}^{\text{EHR}} =&\ell_{\text{orth}}(\mathbf{h}_{\text{shared}}^{\text{CXR}}, \mathbf{h}_{\text{distinct}}^{\text{CXR}}),
\end{aligned}
    \label{eq:disentanglement}
\end{equation}
where
$
    \ell_{\text{orth}} (\mathbf{h}_1, \mathbf{h}_2) = \frac{\left|\langle\mathbf{h}_1, \mathbf{h}_2\rangle\right|}{||\mathbf{h}_1||_2\cdot|| \mathbf{h}_2||_2}
$,
and $\langle\mathbf{h}_1\cdot\mathbf{h}_2\rangle$ denotes the inner product between vectors $\mathbf{h}_1$ and $\mathbf{h}_2$.

\subsection{Disease-aware Masked Attention Fusion}
Inspired by the fact that clinicians rely on different diagnostic tools on varying scales according to the patient's health condition and the particular disease, we propose to learn the significance of each modal regarding predicting different diseases for different patients. To this end, we develop a disease-aware masked attention fusion module that could respect the importance of each modality for different prediction targets.

First, we compute the query vector by taking the average of the available representations following by a linear projection, given by:
\begin{equation}\small
    \mathbf{q} = \begin{cases} (\mathbf{h}^{\text{EHR}}_{\text{distinct}} + \mathbf{h}_{\text{shared}} + \mathbf{h}^{\text{CXR}}_{\text{distinct}})\mathbf{W}^Q/3 &\text{if $\mathbf{X}^{\text{CXR}}\not=\emptyset$},\\
    (\mathbf{h}^{\text{EHR}}_{\text{distinct}} + \mathbf{h}_{\text{shared}})\mathbf{W}^Q/2 &\text{otherwise},
    \end{cases}
\end{equation}

The query vector can be regarded as a summary of the medical status of the patient. To allow different modal significance to be captured, we compute a set of ``target vectors'', each corresponding to a particular prediction target:
\begin{equation}
    \mathbf{K}_c = \mathbf{H}\mathbf{W}_c^K,
\end{equation}
where $c$ denotes the index of the prediction target and $\mathbf{H}$ is obtained by stacking the representations row-wisely: $$\mathbf{H}=\left[(\mathbf{h}^{\text{EHR}}_{\text{distinct}})^\top,~(\mathbf{h}_{\text{shared}})^\top,~ (\mathbf{h}^{\text{CXR}}_{\text{distinct}})^\top\right]\in\mathbb{R}^{3\times d}$$

We follow the scaled-product attention~\cite{vaswani2017attention} to generate the attention weightings of the three representations for each prediction target:
\begin{equation}
    \boldsymbol{\alpha}_c = \operatorname{softmax}\left(\frac{\mathbf{q}\mathbf{K}_c+\mathbf{m}}{\sqrt{d}}\right),\quad c=1,\dots,|C|,
\end{equation}
where $|C|$ is the number of prediction classes and $\mathbf{m}\in\{1, -\infty\}^3$ is a masking vector. It takes value of ones except the third entry, $m_3$, which equals negative infinity when CXR is missing, one otherwise. The final representation for the $c^{\text{th}}$ prediction target is given by:
\begin{equation}
    \tilde{\mathbf{h}}_c = \boldsymbol{\alpha}_c^\top\mathbf{H}\mathbf{W}^V,
\end{equation}
where $\mathbf{W}^Q$, $\mathbf{W}^K_c$, and $\mathbf{W}^V$ are projection matrices.

\subsubsection{Attention ranking loss.} To further enforce the modal significance to be explicitly captured, we propose an attention ranking loss. First, we train auxiliary classifiers using $\mathbf{h}^{\text{EHR}}_{\text{distinct}}$, $\mathbf{h}_{\text{shared}}$, and $\mathbf{h}^{\text{CXR}}_{\text{distinct}}$ as input jointly with the model learning, producing three predictions:
\begin{equation*}\small
    \hat{\mathbf{y}}_1 = g_1(\mathbf{h}^{\text{EHR}}_{\text{distinct}}),~
    \hat{\mathbf{y}}_2 = g_2(\mathbf{h}_{\text{shared}}),
    \text{ and }\hat{\mathbf{y}}_3 = g_3(\mathbf{h}^{\text{CXR}}_{\text{distinct}}),
\end{equation*}
where $g$'s are parameterized by two-layer feedforward networks. We use cross-entropy as the auxiliary loss functions:
\begin{equation}
\begin{aligned}
    &\mathcal{L}_{\text{aux}}=\sum_{i=1}^3 \sum_{c=1}^{|C|} \ell_{ci}\\
    \text{with ~~} &\ell_{ci}= y_c\log(\hat{y}_{ci}) + (1-y_c) \log(1-\hat{y}_{ci})
\end{aligned}
\label{eq:auxiliary_loss}
\end{equation}

The auxiliary loss function reflects the capability of each representation of predicting the target. Thus, we enforce the attention weights to have a ranking consistent with the order of the three loss values. We use the margin ranking loss given by:
\begin{equation}\small
    \mathcal{L}_{\text{attn}}=\frac{1}{2|C|}\sum_{c=1}^{|C|}\sum_{i=1}^3\sum_{j\neq i} \max\big(0, \mathds{1}[\ell_{ci}<\ell_{cj}](\alpha_{cj}-\alpha_{ci})+\epsilon\big),
    \label{eq:attn_loss}
\end{equation}
where $\mathds{1}[\cdot]$ is the indicator function. It equals one if the condition holds, zero otherwise. Eq.~\eqref{eq:attn_loss} imposes penalty when the prediction $y_{ci}$ is better than $y_{cj}$ but the attention weighting $\alpha_{ci}$ is not greater than $\alpha_{cj}$ with a margin of $\epsilon$.

\subsection{Learning Algorithms}
After obtaining the final representations $\tilde{\mathbf{h}}_c$, the final prediction for the $c^{\text{th}}$ class can be obtained using a feedforward layer: $\hat{y}_c=\psi_c(\tilde{\mathbf{h}}_c)$. The loss function for the final prediction can be given by cross entropy as:
\begin{equation}
    \mathcal{L}_{\text{pred}} = \sum_{c=1}^{|C|} y_c\log(\hat{y}_c)+(1-y_c)\log(1-\hat{y}_c).
    \label{eq:final_prediction}
\end{equation}

The overall loss function to minimize is then given by adding the distribution alignment loss in Eq.~\eqref{eq:distribution_alignment}, the disentanglement loss in Eq.~\eqref{eq:disentanglement}, the auxiliary loss in Eq.~\eqref{eq:auxiliary_loss}, the attention ranking loss in Eq.~\eqref{eq:attn_loss}, and the final prediction loss in  Eq.~\eqref{eq:final_prediction}:
\begin{equation}\small
\mathcal{L}=\mathcal{L}_{\text{pred}}+\lambda_1\mathcal{L}_{\text{JSD}}+\lambda_2(\mathcal{L}_{\text{orth}}^{\text{EHR}}+\mathcal{L}_{\text{orth}}^{\text{CXR}})+\lambda_3(\mathcal{L}_{\text{attn}}+\mathcal{L}_{\text{aux}}).
\label{eq:overall_loss}
\end{equation}

\paragraph{Training with missing modality.} %
When CXR is not available, we extract the disentangled representation from EHR data only and use the EHR shared representation as $\mathbf{h}_{\text{shared}}$ directly, as in Eq.~\eqref{eq:pool_shared}, and loss terms in Eq.~\eqref{eq:overall_loss} involving CXR representations are removed. Therefore, the objective function to be optimized over the entire training set with partially missing CXR data is given by:
\begin{equation}
    \min~~\frac{1}{|\mathcal{D}|}\left(\sum_{i\in\mathcal{D}_{\text{paired}}} \mathcal{L}_i + \sum_{i\in\mathcal{D}_{\text{partial}}} \left(\mathcal{L}_{\text{pred}}+\lambda_2\mathcal{L}_{\text{orth}}^{\text{EHR}}\right)\right),
\end{equation}
where $i$ is the index of patients.

\begin{figure}
    \centering
    \includegraphics[width=0.85\linewidth]{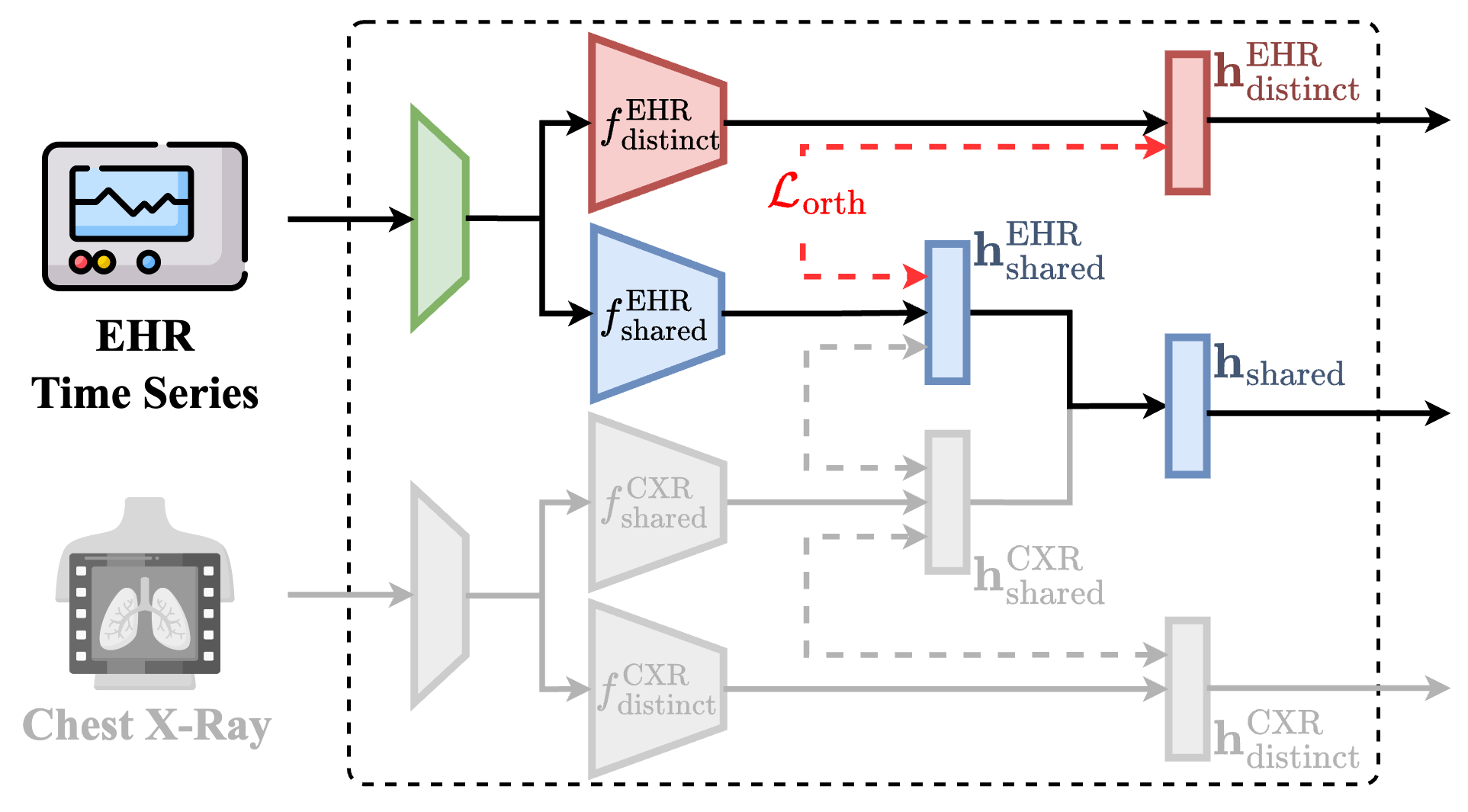}
    \caption{Data flow in the disentangled representation learning module when the CXR modality is missing. The shared representation extracted from EHR will be directly used as $\mathbf{h}_{\text{shared}}$. Inactive components and loss terms are grayed out.}\vspace{-1em}
    \label{fig:missing_cxr}
\end{figure}

\section{Experiments}

\subsection{Experiment settings}
\subsubsection{Datasets and preprocessing.}
We use the large-scale real-world EHR datasets, MIMIC-IV~\cite{johnson2023mimic} and MIMIC-CXR~\cite{johnson2019mimic} to empirically evaluate the predictive performance of \drfuse. MIMIC-IV contains de-identified data of adult patients admitted to either intensive care units or the emergency department of Beth Israel Deaconess Medical Center (BIDMC) between 2008 and 2019. MIMIC-CXR is a publicly available dataset of chest radiographs collected from BIDMC, where a subset of patients can be matched with those in MIMIC-IV.

We follow similar procedures to preprocess the data as those in~\cite{mlhc2022hayatmedfuse}. We extract 17 clinical variables that are routinely monitored in ICU, including five categorical variables
and twelve continuous ones. %
We use the disease prediction as the prediction task, where the 25 disease phenotype labels are generated based on diagnosis codes following~\cite{harutyunyan2019multitask}. To better align with the clinical need for early prediction, we make predictions of the disease phenotypes using data within the first 48 hours of the ICU admission. Accordingly, we retrieve the last Anterior-Posterior(PA) projection chest X-ray in the same observation window. In total, we extracted $59,344$ ICU stays with EHR records, of which $10,630$ are associated with CXR.

\begin{table}\small
    \centering
    \setlength{\tabcolsep}{4pt}
    \begin{tabular}{c c c c c}
    \toprule
    & {\begin{tabular}[c]{@{}c@{}}Missing\\ Modality\end{tabular}} & Training & Validating & Testing\\
    \midrule
    \textit{full dataset} & \checkmark & 42,628 & 4,802 &11,914 \\
    \textit{matched subset} &$\times$& 7,637 & 857 & 2,136  \\
    \bottomrule
    \end{tabular}
    \caption{Number of samples in the two datasets constructed.}
    \label{tab:data_distribution}\vspace{-1em}
\end{table}

\begin{table*}[h]
\small
\centering
\begin{tabular}{lcccc}
\toprule
\multicolumn{1}{c}{} & \multicolumn{2}{c}{{Trained with the \textit{matched subset}}} & \multicolumn{2}{c}{{Trained with the \textit{full dataset}}} \\\cmidrule(lr){2-3}\cmidrule(lr){4-5}
\multicolumn{1}{c}{\multirow{-2}{*}{{Model}}} & {testing on \textit{matched subset}}
 & {testing on \textit{full dataset}}   & {testing on \textit{matched subset}}
 & {testing on \textit{full dataset}} \\\midrule
\textbf{Transformer} & 0.408 (0.368, 0.455) & 0.374 (0.355, 0.395) & 0.435 (0.393, 0.481) & 0.418 (0.398, 0.440) \\
\textbf{MMTM} & 0.416 (0.378, 0.462)          & 0.359 (0.342, 0.379)          & 0.422 (0.383, 0.469)          & 0.407 (0.387, 0.428)\\
\textbf{DAFT} & 0.417 (0.376, 0.462)          & 0.348 (0.331, 0.368)          & 0.430 (0.389, 0.477)          & 0.409 (0.389, 0.431)\\
\textbf{MedFuse} & 0.427 (0.387, 0.473)          & 0.329 (0.312, 0.347)          & 0.434 (0.394, 0.481)          & 0.405 (0.385, 0.427) \\
\textbf{MedFuse-II}                    & 0.418 (0.378, 0.463)                                & 0.329 (0.314, 0.348)                               & 0.427 (0.387, 0.473)                            & 0.412 (0.391, 0.433)                                \\
\textbf{DrFuse} & \textbf{0.450 (0.426, 0.498)} & \textbf{0.384 (0.371, 0.402)} & \textbf{0.470 (0.420, 0.512)} & \textbf{0.419 (0.391, 0.434)} \\\bottomrule
\end{tabular}
\caption{Overall performance measured by the macro average of PRAUC over all 25 disease phenotype labels for different combinations of training and test subset. Numbers in bold indicates the best performance in each column. \drfuse consistently outperforms all baselines in all settings with a significant margin.}\label{tab:overall_performance}\vspace{-1em}
\end{table*}

\begin{table*}[ht]
\small
\centering
\setlength{\tabcolsep}{4pt}
\begin{tabular}{@{}r|c|cc|lll@{}}
\toprule
\multicolumn{1}{c|}{\textbf{Disease Label}}               & \textbf{Prevalence} & \textbf{\begin{tabular}[c]{@{}c@{}}ResNet50\\ (CXR)\end{tabular}} & \textbf{\begin{tabular}[c]{@{}c@{}}Transformer\\ (EHR)\end{tabular}} & \multicolumn{1}{c}{\textbf{MedFuse}} & \multicolumn{1}{c}{\textbf{MedFuse-II}} & \multicolumn{1}{c}{\textbf{DrFuse}} \\ \midrule
Acute and unspecified renal failure                & 0.32                & 0.469                                                             & 0.537                                                                & 0.559 (4.1\%)                          & 0.541 (0.7\%)                          & 0.541 (0.7\%)                         \\
Acute cerebrovascular disease                      & 0.07                & 0.145                                                             & 0.457                                                                & 0.461 (0.9\%)                          & 0.441 (-3.5\%)                         & 0.441 (-3.5\%)                        \\
Acute myocardial infarction                        & 0.09                & 0.165                                                             & 0.170                                                                & {\color[HTML]{0002FF} 0.217 (27.6\%)}  & 0.177 (4.1\%)                          & {\color[HTML]{0002FF} 0.193 (13.5\%)} \\
Cardiac dysrhythmias                               & 0.38                & 0.566                                                             & 0.513                                                                & {\color[HTML]{FE0000} 0.552 (-2.5\%)}  & {\color[HTML]{FE0000} 0.517 (-8.7\%)}  & 0.568 (0.4\%)                         \\
Chronic kidney disease                             & 0.24                & 0.400                                                             & 0.424                                                                & {\color[HTML]{0002FF} 0.455 (7.3\%)}   & {\color[HTML]{0002FF} 0.455 (7.3\%)}   & {\color[HTML]{0002FF} 0.445 (5\%)}    \\
Chronic obstructive pulmonary disease              & 0.15                & 0.374                                                             & 0.239                                                                & {\color[HTML]{FE0000} 0.323 (-13.6\%)} & {\color[HTML]{FE0000} 0.317 (-15.2\%)} & {\color[HTML]{FE0000} 0.355 (-5.1\%)} \\
Complications of surgical/medical care             & 0.22                & 0.303                                                             & 0.408                                                                & {\color[HTML]{0002FF} 0.379 (-7.1\%)}  & 0.395 (-3.2\%)                         & 0.407 (-0.2\%)                        \\
Conduction disorders                               & 0.11                & 0.625                                                             & 0.237                                                                & {\color[HTML]{FE0000} 0.372 (-40.5\%)} & {\color[HTML]{FE0000} 0.231 (-63\%)}   & 0.619 (-1\%)                          \\
Congestive heart failure; nonhypertensive          & 0.29                & 0.593                                                             & 0.509                                                                & 0.597 (0.7\%)                          & {\color[HTML]{FE0000} 0.558 (-5.9\%)}  & {\color[HTML]{0002FF} 0.629 (6.1\%)}  \\
Coronary atherosclerosis and related               & 0.34                & 0.657                                                             & 0.559                                                                & {\color[HTML]{FE0000} 0.603 (-8.2\%)}  & {\color[HTML]{FE0000} 0.588 (-10.5\%)} & 0.640 (-2.6\%)                        \\
Diabetes mellitus with complications               & 0.12                & 0.217                                                             & 0.520                                                                & {\color[HTML]{FE0000} 0.469 (-9.8\%)}  & 0.505 (-2.9\%)                         & {\color[HTML]{FE0000} 0.486 (-6.5\%)} \\
Diabetes mellitus without complication             & 0.21                & 0.276                                                             & 0.361                                                                & {\color[HTML]{FE0000} 0.338 (-6.4\%)}  & 0.363 (0.6\%)                          & {\color[HTML]{0002FF} 0.381 (5.5\%)}  \\
Disorders of lipid metabolism                      & 0.41                & 0.587                                                             & 0.593                                                                & 0.598 (0.8\%)                          & 0.612 (3.2\%)                          & 0.612 (3.2\%)                         \\
Essential hypertension                             & 0.44                & 0.558                                                             & 0.578                                                                & 0.592 (2.4\%)                          & 0.601 (4\%)                            & 0.572 (-1\%)                          \\
Fluid and electrolyte disorders                    & 0.45                & 0.563                                                             & 0.675                                                                & 0.675 (0\%)                            & 0.663 (-1.8\%)                         & 0.660 (-2.2\%)                        \\
Gastrointestinal hemorrhage                        & 0.07                & 0.121                                                             & 0.193                                                                & {\color[HTML]{FE0000} 0.152 (-21.2\%)} & {\color[HTML]{FE0000} 0.152 (-21.2\%)} & {\color[HTML]{0002FF} 0.204 (5.7\%)}  \\
Hypertension with complications                    & 0.22                & 0.378                                                             & 0.393                                                                & {\color[HTML]{0002FF} 0.418 (6.4\%)}   & {\color[HTML]{0002FF} 0.424 (7.9\%)}   & 0.409 (4.1\%)                         \\
Other liver diseases                               & 0.17                & 0.341                                                             & 0.268                                                                & 0.351 (2.9\%)                          & 0.319 (-6.5\%)                         & {\color[HTML]{0002FF} 0.389 (14.1\%)} \\
Other lower respiratory disease                    & 0.13                & 0.182                                                             & 0.170                                                                & 0.167 (-8.2\%)                         & 0.176 (-3.3\%)                         & 0.186 (2.2\%)                         \\
Other upper respiratory disease                    & 0.05                & 0.102                                                             & 0.165                                                                & {\color[HTML]{FE0000} 0.114 (-30.9\%)} & 0.161 (-2.4\%)                         & {\color[HTML]{0002FF} 0.205 (24.2\%)} \\
Pleurisy; pneumothorax; pulmonary collapse         & 0.10                & 0.195                                                             & 0.126                                                                & 0.191 (-2.1\%)                         & {\color[HTML]{FE0000} 0.156 (-20\%)}   & 0.192 (-1.5\%)                        \\
Pneumonia                                          & 0.18                & 0.354                                                             & 0.404                                                                & 0.400 (-1\%)                           & {\color[HTML]{0002FF} 0.428 (5.9\%)}   & 0.419 (3.7\%)                         \\
Respiratory failure; insufficiency; arrest (adult) & 0.28                & 0.520                                                             & 0.607                                                                & 0.591 (-2.6\%)                         & 0.605 (-0.3\%)                         & 0.615 (1.3\%)                         \\
Septicemia (except in labor)                       & 0.22                & 0.371                                                             & 0.538                                                                & 0.522 (-3\%)                           & 0.514 (-4.5\%)                         & 0.528 (-1.9\%)                        \\
Shock                                              & 0.17                & 0.342                                                             & 0.558                                                                & 0.567 (1.6\%)                          & 0.545 (-2.3\%)                         & 0.542 (-2.9\%)                        \\ \midrule
\multicolumn{2}{r|}{Average Rank} & 3.96 & 3.24 & \multicolumn{1}{c}{2.68} & \multicolumn{1}{c}{2.84} & \multicolumn{1}{c}{\textbf{2.04}} \\\bottomrule
\end{tabular}
\caption{The PRAUC score for each disease label. ``ResNet50'' and ``Transformer'' indicate the performance obtained using only CXR data and EHR data, respectively. The percentages within parentheses indicate the relative difference against the best uni-modal prediction. Results show that \drfuse could better address the inconsistency issue, resulting in the highest average rank over all disease labels. Results with relative differences beyond $\pm$5\% over the best uni-modal predictions are highlighted. }\label{tab:performance_per_disease}\vspace{-0.5em}
\end{table*}

To test \drfuse in different modality missing settings, we construct two datasets using the extracted data: a \textit{full dataset} containing all patients regardless of having CXR or not, and a \textit{matched subset} only containing patients having both EHR and CXR. We randomly split the dataset with a ratio of $7$:$1$:$2$ for training, validation, and testing. It is worth noting that patients in the validation and test subsets of the \textit{matched subset} are also split into validation and test subsets, respectively. This allowed us to train the model using one dataset and test it with the other. Table \ref{tab:data_distribution} shows the number of patients having each data modality.

\subsubsection{Evaluation metrics.} Due to the highly imbalanced nature of the disease labels (see Table \ref{tab:performance_per_disease} for prevalence), we evaluate the performance of \drfuse and baseline models using Area Under the Precision Recall Curve (PRAUC).

\paragraph{Experiment implementation.} The experiment environment is a machine equipped with dual Intel Xeon Silver 4114 CPUs and four Nvidia V100 GPU cards. The model is implemented based on Pytorch 2.0.1. We use grid search to tune the hyperparameters using the validation set and report that over the test set. The search spaces of the hyperparameters are: $\lambda_1\in\{0,0.1,\mathbf{1}\}$, $\lambda_2\in\{0, 0.1, \mathbf{1}\}$, $\lambda_3\in\{0, \mathbf{0.5}, 1\}$, $\text{lr}\in\{\mathbf{0.0001}, 0.001\}$, where the value in bold indicates the optimal choice. When training with the matched subset, we randomly remove CXR of 30\% samples within each mini-batch as an additional data augmentation.

\subsection{Baseline Models}
We compare against the following baselines.
\begin{itemize}
    \item \textbf{MMTM}~\cite{joze2020mmtm} is a module that can leverage the information between modalities with flexible plug-in architectures. %
    Since the model assumes full modality, we compensate the missing modality CXR with all zeros during training and testing.
    \item \textbf{DAFT}~\cite{daft-polsterl2021combining} is a module that can be plugged into CNN models to achieve information exchange between tabular data and image modality. Similarly, we replace the input of CXR with matrices of all zeros during training and testing.
    \item \textbf{MedFuse}~\cite{mlhc2022hayatmedfuse} uses an LSTM-based fusion to combine features from the image encoder and EHR encoder. Missing modality is handled by learning a global representation for the missing CXR.
    \item \textbf{MedFuse-II} is a variant of MedFuse with its CXR encoders and EHR encoders replaced by ResNet50 and Transformer, respectively, to ensure fair comparison with \drfuse.
    \item \textbf{Transformer}~\cite{vaswani2017attention} is the EHR encoder used by \drfuse, which is a uni-modal method that takes only EHR as input.
\end{itemize}

\subsection{Overall Performance of Disease Prediction}
The performance in terms of disease phenotype prediction is summarized in Table~\ref{tab:overall_performance}. We report the macro average of PRAUC over all 25 disease phenotype labels together with the corresponding 95\% confidence interval obtained through 1000 iterations using the bootstrap method. The results show that \drfuse consistently outperforms all baselines compared with a large margin. When trained and tested both with the \textit{matched subset}, \ie, no missing modality is involved, \drfuse achieves 5.4\% relative improvement against MedFuse, demonstrating that the proposed \drfuse could achieve effective modality fusion. When trained with the \textit{full dataset} and tested with the \textit{matched subset}, \drfuse achieves 8\% relative improvement against MedFuse, suggesting that \drfuse could fully utilize the training samples with missing modalities. When tested on the \textit{full dataset}, all methods, including the uni-modal Transformer, obtain worse results comparing with the test scores obtained on the \textit{matched subset}. This suggests that severe domain shift could exist between the two subsets. This might be because patients who could not undergo X-ray scans may have more complex health conditions and thus are much harder to predict. Having said so, \drfuse still obtains the best performance in the presence of such potential severe domain shift in the \textit{full dataset} benefited from the representation disentanglement.

\begin{figure}
    \centering
    \includegraphics[width=0.88\columnwidth]{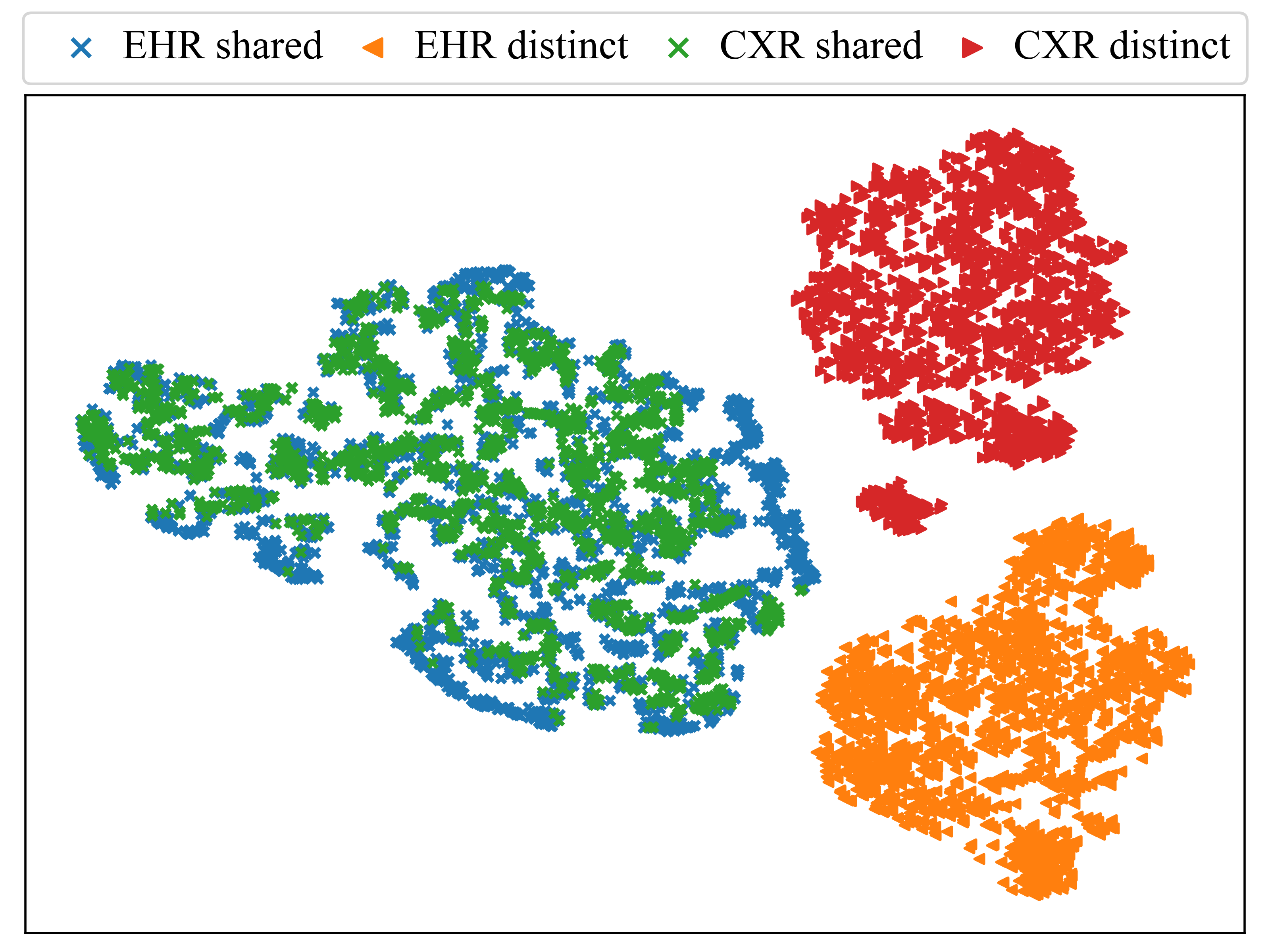}
    \caption{t-SNE visualization of distinct and shared features for the test set in the \textit{matched subset}. \drfuse could well align the distributions of the EHR and CXR shared representations, as well as disentangling the distinct representations.}
    \label{fig:disentangle}\vspace{-1em}
\end{figure}

\subsection{Disease-Wise Prediction Performance}
To gain more insights into the prediction performance, we show the disease-wise PRAUC scores obtained by the uni-modal methods, MedFuse, and \drfuse in Table~\ref{tab:performance_per_disease}. Numbers inside parentheses indicate the relative difference against the best uni-modal prediction. The results show that combining EHR and CXR is not always helpful for all diseases, due to the modal inconsistency issue as mentioned earlier. For example, when predicting \textit{conduction disorders} and \textit{other upper respiratory disease}, the performance of MedFuse drops 40.5\% and 30.9\%, respectively, compared with uni-modal predictions. On the contrary, \drfuse only drops 1\% for \textit{conduction disorders} and achieves 24.2\% improvement for \textit{other upper respiratory disease}. This demonstrates that the proposed \drfuse could better address the modal inconsistency issue by inferring the disease-specific and patient-specific modal significance.

\subsection{Visualization of Disentangled Representation}

To further validate the effectiveness of the disentangled representation learning, we visualize the shared and distinct representations for EHR and CXR data with t-SNE in Fig.~\ref{fig:disentangle}. The shared representations, $\mathbf{h}_{\text{shared}}^{\text{EHR}}$ and $\mathbf{h}_{\text{shared}}^{\text{CXR}}$, are well blended as a cluster. Meanwhile the distinct representations, $\mathbf{h}_{\text{distinct}}^{\text{EHR}}$ and $\mathbf{h}_{\text{distinct}}^{\text{CXR}}$, remain well-separated not just from each other but also from the shared features.  %

\begin{table}\small
    \centering
    \setlength{\tabcolsep}{4pt}
    \begin{tabular}{rcc}
    \toprule
    Model & \begin{tabular}[c]{@{}c@{}}PRAUC\\ @\textit{matched subset}\end{tabular}& \begin{tabular}[c]{@{}c@{}}PRAUC\\ @\textit{full dataset}\end{tabular} \\\midrule
         w/o disentangled & 0.446 (0.411, 0.501) & 0.374 (0.355, 0.395) \\
         MSE alignment & 0.447 (0.410, 0.498) & 0.375 (0.356, 0.396)\\
         w/o attn. ranking & 0.438 (0.396, 0.485) & 0.361 (0.343, 0.382) \\\midrule
    \drfuse & 0.450 (0.426, 0.498) & 0.384 (0.371, 0.402) \\\bottomrule
    \end{tabular}
    \caption{Results of the ablation study tested over different datasets by removing each component from \drfuse. The models are trained using the \textit{matched subset}.}
    \label{tab:ablation}\vspace{-1em}
\end{table}

\subsection{Ablation Study}
To gain further insights to the source of performance gain of \drfuse, we conduct ablation study by training the model using the \textit{matched subset} with each component of \drfuse removed. The results are summarized in Table~\ref{tab:ablation}. The first row is obtained by removing $\mathcal{L}_{\text{JSD}}$ and $\mathcal{L}_{\text{orth}}$ and the second row is obtained by replacing the JSD with the MSE loss and the logit pooling with the average pooling. A significant performance drop can be observed when tested using the \textit{full dataset} with missing modality. This demonstrates that the proposed disentangled representation learning is effective in handling missing modality. The third row removes the attention ranking loss $\mathcal{L}_{\text{attn}}$, where significant performance drop is obtained, showing that capturing disease-wise modal significance is important for the disease prediction task and the proposed method is effective in achieving this goal.

\vspace{-0.5em}\section{Conclusion}
In this paper, we propose a novel model, \drfuse, that learns the disentangled representation from EHR and CXR data to achieve medical multi-modal data fusion in the presence of missing modality and modal inconsistency. A shared representation and a distinct representation are learned from each modal. We align the shared representations via minimizing the Jensen–Shannon divergence (JSD) and achieve representation disentanglement via imposing orthogonal constraints. A logit pooling operation is derived to fuse the shared representations. Besides, we propose a disease-aware attention fusion module that captures the patient-specific modal significance for each prediction target via an attention ranking loss. The experimental results demonstrate that the proposed model is effective in achieving disentangled representation, addressing the missing modality and modal inconsistency issues; thus achieving significant performance improvement. For future research directions, we will focus on addressing the domain shift between patient with and without CXR jointly with the multi-modal learning.

\section*{Acknowledgements}
The work described in this paper is supported by a grant of Hong Kong RGC Theme-based Research Scheme (project no. T45--401/22--N), an Innovation and Technology Fund--Midstream Research Programme for Universities (ITF--MRP) (project no. MRP/022/20X), and General Research Fund RGC/HKBU12201219 from the Research Grant Council.
\bibliography{aaai24}

\end{document}